\documentclass[12pt]{article}
\usepackage{graphicx}
\usepackage{slashed}
\begin{document}

\title{Infinite Classes of Cases with Non-trivial Anomaly Cancellation}
\author{{\bf S.M. Barr} \\
Department of Physics and Astronomy \\ 
and \\
Bartol Research Institute \\ University of Delaware
Newark, Delaware 19716} \maketitle

\begin{abstract}
It is pointed out that there are infinite classes of cases based on gauge groups of the form $SU(p) \times SU(q) \times U(1)$ in which gauge anomalies cancel non-trivially for small sets of fermion multiplets that include symmetric tensor representations.
These cancellations are non-trivial in the sense that no group-theoretic explanation in terms of embedding in a larger simple group is apparent. The cases presented here could be useful for model building and lead to models with extra leptons and an extra $U(1)$ gauge interaction under which the Standard Model
fermions have distinctive charges. In many cases the $SU(q) \times U(1)$ groups act as family symmetries. 
\end{abstract}

\section{Introduction}

Many models of particle interactions make use of products of gauge 
groups. Some well-known examples are the $SU(3) \times SU(2) \times U(1)$ of the
Standard Model, the $SU(4) \times SU(2) \times SU(2)$ of the Pati-Salam
model \cite{PatiSalam}, the $SU(5) \times U(1)$ of Flipped SU(5) models \cite{FlippedSU(5)}, and the $SU(3) \times SU(3) \times SU(3)$ of 
trinification models \cite{trinification}. Obviously,
it is necessary that the set of fermion multiplets of a model be 
chiral under the gauge group and that all gauge
anomalies cancel. In general, this is a highly restrictive constraint,
especially if there are several groups and only a few fermion multiplets.

Consider for example, the fermion multiplets of one family in Flipped
$SU(5)$. Under the $SU(5) \times U(1)$ they transform as
${\bf 10}^1 + \overline{{\bf 5}}^{-3} + {\bf 1}^5$, where the superscripts denote
the $U(1)$ charges. It seems surprising that there exist $U(1)$
charge assignments for these three $SU(5)$ multiplets that satisfy all
anomaly conditions, for that
requires two numbers to satisfy three relations. The two numbers are
the ratios of the three charges (the overall normalization of $U(1)$
charges does not matter), and the three relations are the 
$SU(5)^2 U(1)$, $U(1)$-gravity, and $U(1)^3$ anomaly conditions. Moreover, 
realistic models have charges that are relatively rational (in order
to allow the required Yukawa couplings and Higgs self-couplings to exist), and 
one of the anomaly conditions on the $U(1)$ 
charges is a cubic equation whose solutions would typically be irrational
numbers. And yet the charges $1$, $-3$, and $5$ do indeed satisfy all the conditions.
$SU(5)^2 U(1)$ anomaly:  $3(1) + 1(-3) + 0(5) = 0$; 
$U(1)$-gravity anomaly:  $10(1) + 5(-3) + 1(5) = 10 -15 +5 =0$; and 
$U(1)^3$ anomaly:  $10(1)^3 + 5(-3)^3 + 1(5)^3 = 10 - 135 + 125=0$.

\noindent What at first seems like a strange coincidence becomes less
surprising when it is realized that there is a group-theoretical
explanation for it. The group $SU(5) \times U(1)$ is a subgroup
of $SO(10)$, and when the ${\bf 16}$ (spinor) multiplet of
$SO(10)$ is decomposed under this subgroup, it yields exactly 
${\bf 10}^1 + \overline{{\bf 5}}^{-3} + {\bf 1}$.  
The anomaly-freedom of $SO(10)$, which is easily proven, thus
explains the cancellation that appears surprising at the 
level of $SU(5) \times U(1)$.

This example is typical. When anomaly cancellation appears ``surprising",
in the sense that more anomaly conditions are satisfied than simple 
counting of equations and unknowns would suggest, it 
almost always turns out that there is a group-theoretical
explanation in terms of embedding in a larger simple group. An interesting
case where a group-theoretic explanation for anomaly cancellation
exists but is somewhat hidden can be found in \cite{peculiarU(1)}. 

In this paper, we point out the existence of several infinite
classes of cases in which the anomaly cancellation seems quite surprising 
in the above sense and yet no group-theoretic explanation is apparent. 
This is what me mean by the ``non-trivial" cancellation of anomalies referred to in the title of this paper. In section 2, we present an infinite class with rank-2 tensor fermions and gauge group $SU(N) \times SU(N-4) \times U(1)$. In section 3, we present two infinite classes with rank-3 tensor fermions, one with gauge group $SU(N) \times SU(N-3) \times U(1)$ and the other with gauge group $SU(N) \times SU(N-6) \times U(1)$. We also present two infinite classes with rank-4 tensor fermions, one with gauge group $SU(N) \times SU(N-4) \times U(1)$ and the other with gauge group $SU(N) \times SU(N-8) \times U(1)$. It seems probable that there are an infinite number of such infinite classes
 
\section{A class based on $SU(N) \times SU(N-4) \times U(1)$ with rank-2-tensor fermions}

For this class, the gauge group is $SU(N) \times SU(N-4) \times U(1)$, with $N \geq 5$
and the anomaly-free set of fermions is

\begin{equation}
\{ {\rm fermion \;\; multiplets} \} \;\;\; =  \;\;\; ([2], (0))^{(N-4)}  +  (\overline{[1]}, \overline{(1)})^{-(N-2)}  +  
([0], (2))^N.
\end{equation}

\noindent The notation here is as follows. The superscripts $(N-4)$, $-(N-2)$ and $N$ are the $U(1)$ charges.
$[p]$ stands for an antisymmetric rank-$p$ 
tensor, $\overline{[p]}$ stands for the conjugate of that, and $(p)$ stands for a symmetric rank-$p$ tensor. (Of course, $[0] = (0)$ and $[1] = (1)$.) One could also denote the multiplets by their dimensions, and write this set as
$({\bf \frac{N(N-1)}{2}}, {\bf 1})^{(N-4)} + (\overline{{\bf N}}, \overline{{\bf (N-4)}})^{-(N-2)} + ({\bf 1}, {\bf \frac{N(N+1)}{2}})^N$. Or one could show indices explicitly and write it as $(\psi^{[\alpha \beta]})^{(N-4)} + (\psi_{\alpha, m})^{-(N-2)} + (\psi^{(mn)})^{N}$, where $SU(N)$ indices are Greek and $SU(N-4)$ indices are Latin.
 
Note that the cancellation of the anomalies for this set is highly non-trivial.
First, the cancellation of the $SU(N)^3$ anomaly and of the $SU(N-4)^3$ anomaly requires that two relationships hold between the anomalies of the multiplets of one group and the dimensions of the multiplets of the other. Second, for the cancellation of all anomalies involving the $U(1)$, {\it two} $U(1)$ charge ratios that must satisfy {\it four} anomaly conditions (one of which is cubic in the $U(1)$ charges), namely the $SU(N)^2 U(1)$, $SU(N-4)^2 U(1)$, $U(1)$-gravity, and $U(1)^3$ anomalies. Note the fact that the third fermion multiplet in the set shown in Eq. (1) is a {\it symmetric} rank-2 tensor representation under $SU(N-4)$. It seems very unlikely, therefore, that this set can be obtained from decomposing representations of a larger simple group, since such a decomposition would include a symmetric tensor representation of $SU(N)$. 

Let us denote by $A_p (R)$ and $C_p (R)$ the cubic anomaly and the quadratic Casimir 
coefficients of the representation $R$ of the group $SU(p)$, normalizing so that $A_p([1]) =1$
and $C_p([1]) = 1$. (More precisely, $C_p (R) \equiv tr_R (\lambda^2)/ tr_f (\lambda^2)$, where $\lambda$ is any group generator, and $f$ is the fundamental representation.) Then the dimensions, cubic anomalies and quadratic Casimir coefficients for small symmetric and antisymetric tensor representations are given in Table I. 

\noindent {\bf Table I:} The dimensions, cubic anomalies and quadratic Casimir coefficients of small representations of $SU(p)$
\[
\begin{tabular}{llll}
$R$ & $d(R)$ & $A_p (R)$ & $C_p (R)$ \\
\hline 
$[0] = (0)$ & $1$ & 0 & 0 \\ 
$[1] = (1)$ & $p$ & 1  &  1 \\ 
$[2]$ & $\frac{1}{2} p (p-1)$ & $p-4$ & $p-2$ \\ 
$(2)$ & $\frac{1}{2} p (p+1)$ & $p+4$ & $p+2$ \\ 
$[3]$ & $\frac{1}{6} p (p-1)(p-2)$ & $\frac{1}{2}(p-3)(p-6)$ & $\frac{1}{2}(p-2)(p-3)$ \\ 
$(3)$ & $\frac{1}{6} p (p+1)(p+2)$ & $\frac{1}{2}(p+3)(p+6)$ & $\frac{1}{2}(p+2)(p+3)$ 
\end{tabular}
\]

\noindent Using the numbers in Table I, the cancellation of all anomalies for the set of fermions shown in Eq.(1) is easily verified: 

\begin{equation}
\begin{array}{l}
SU(N)^3 \; {\rm anomaly}  
=(N-4) \cdot (1) \;\; + \;\; (-1) \cdot (N-4) = 0 \\  \\
SU(N-4)^3 \; {\rm anomaly}
= (N) \cdot (1) \;\; + \;\; (1) \cdot (-N) = 0 \\ \\
SU(N)^2 U(1) \; {\rm anomaly} 
= (N-2) \cdot (1) \cdot (N-4) \;\; + \;\; (1) \cdot (N-4) \cdot (2-N) = 0 \\  \\
SU(N-4)^2 U(1) \; {\rm anomaly} 
=  (N) \cdot (1) \cdot (-N+2) \;\; + \;\; (1) \cdot (N-2) \cdot (N) = 0 \\  \\
U(1){\rm -grav} \; {\rm anomaly} = ( \frac{1}{2} N(N-1)) \cdot (1) \cdot (N-4)   
\; + \; (N) \cdot (N-4) \cdot (2-N) \\
\; + \; (1) \cdot \left( \frac{1}{2} (N-4)(N-3) \right) \cdot N = 0 \\  \\
U(1)^3 \; {\rm anomaly} 
= (\frac{1}{2}N(N-1)) \cdot (1) \cdot (N-4)^3  \; + \; (N) \cdot (N-4) \cdot (-N+2)^3 \\
\; + \; (1) \cdot (\frac{1}{2} (N-4)(N-3)) \cdot (N)^3 \\ 
= \frac{1}{2} N(N-4) \left[ (N-1)(N-4)^2 - 2(N-2)^3  + (N-3) N^2 \right] \\
= \frac{1}{2}N (N-4) [(N^3 - 9 N^2 + 24 N -16) -2 (N^3 -6N^2 + 12N -8) + (N^3 -3N^2)] = 0.
\end{array}
\end{equation}

For concreteness, we display in Table II the smallest cases in the class, $N = 5, 6, 7,$ and $8$. For $N =5$, the group 
$SU(N-4)$ does not exist, so this case is degenerate and has only one non-abelian factor in the gauge group.

\noindent {\bf Table II:} The $N=5,6,7,$ and $8$ cases of the infinite class based on $SU(N) \times SU(N-4) \times U(1)$ and rank-2 tensor fermion multiplets.
\[
\begin{tabular}{lll}
{\bf N} & {\bf gauge group} & {\bf fermion multiplets} \\
\hline 
$5$ & $SU(5) \times U(1)$  & ${\bf 10}^1 + \overline{{\bf 5}}^{-3} + {\bf 1}^5$ \\ & & \\
$6$ & $SU(6) \times SU(2) \times U(1)$ & $({\bf 15}, {\bf 1})^1 + (\overline{{\bf 6}}, {\bf 2})^{-2}
+ ({\bf 1}, {\bf 3})^3$ \\ & & \\
$7$ & $SU(7) \times SU(3) \times U(1)$ & $({\bf 21}, {\bf 1})^3 + (\overline{{\bf 7}}, \overline{{\bf 3}})^{-5}
+ ({\bf 1}, {\bf 6})^7$ \\ & & \\
$8$ & $SU(8) \times SU(4) \times U(1)$ & $({\bf 28}, {\bf 1})^2 + (\overline{{\bf 8}}, \overline{{\bf 4}})^{-3}
+ ({\bf 1}, {\bf 10})^4$ 
\end{tabular}
\]

\noindent For the cases of even $N$, all the $U(1)$ charges given by the general expression in Eq. (1) are even, so they are divided by 2 in Table I. 

The $N=5$ case is the well-known one, discussed above, that can be embedded in $SO(10)$. (It is the only case that does not have a symmetric rank-2 tensor multiplet.) The $N=6$ case contains $({\bf 15}, {\bf 1}) + (\overline{{\bf 6}}, {\bf 2})$ of $SU(6) \times SU(2)$, which often appears in model-building \cite{SU(6)xSU(2)} and can be embedded in $E_6$. But here there is an extra $U(1)$ factor, making the total rank of the group 7, meaning that the $N=6$ case is obviously not embeddable in $E_6$. Moreover, even without the extra $U(1)$, the extra $({\bf 1}, {\bf 3})$ of fermions would mean that it was not embeddable in $E_6$ or any other simple group, because such a symmetric tensor representation would be accompanied by symmetric tensors of $SU(6)$.
And for the same reason, as mentioned above, no other case (except the degenerate one $N=5$) is embeddable in a larger simple group.

On the other hand, every case in the infinite class is contained in all the cases with larger $N$ and can arise from them through spontaneous symmetry breaking by the VEV of a Higgs fields in the
bi-fundamental representation $([1], \overline{[1]})$. For example, in the case $N=7$, a Higgs field in the representation $({\bf 7}, \overline{{\bf 3}})$ can both break the group down to $SU(6) \times SU(2) \times U(1)$ and give mass to the extra fermions, leaving just the fermions of the $N=6$ case with the correct $U(1)$ charges (as must be so, of course, because of the anomaly cancellation).
For every case in the class, the set of fermions shown in Eq. (1) contains exactly one chiral family when decomposed under the Standard Model group, along with some vectorlike multiplets and some extra leptons coming from the symmetric rank-2 tensor multiplet $([0], (2))$. 

It is well-known that for an $SU(N)$ grand unified theory, the simplest anomaly-free set of fermion multiplets that contains one Standard Model family is the rank-2 antisymmetric tensor plus $N-4$ copies of the anti-fundamental representation ({\it i.e.} a $[2]$ plus $N-4$ copies of $\overline{[1]}$). What is being done here is to make those $N-4$ copies into a multiplet of an additional $SU(N-4)$ gauge group. That requires cancelling the anomaly of the $SU(N-4)$. It just happens that this can be done with just a $(2)$ ({\it i.e.} the rank-2 symmetric tensor) of $SU(N-4)$. The coincidence here is that the anomaly of a $[2]$  of $SU(N)$ is $N-4$, while the anomaly of the $(2)$ of $SU(N-4)$ is $N$.
The further remarkable fact is that with just these $SU(N) \times SU(N-4)$ fermion multiplets a further $U(1)$ gauge symmetry can be introduced under which the fermions are charged and anomaly free. As already mentioned, that involves two ratios of charges satisfying four anomaly cancellation relations.  

For every case, the $U(1)$ charges are such as to allow flipped embeddings of 
the Standard Model in the $SU(N)$ group. This actually follows trivially 
from the fact that the $N=5$ case is embedded in every higher case. 
It is also noteworthy that there are two ways of embedding the electroweak
group $SU(2)_L$: for each $N>5$, it can be a subgroup of the $SU(N)$ factor
(along with the color group) or of the $SU(N-4)$ factor. In the former 
case, the extra leptons in the symmetric rank-2 tensor are neutral
under the electroweak group. In the latter case, they are in 
multiplets of $SU(2)_L$. In some cases these extra leptons can be given
vectorlike masses that would allow their masses to be much 
larger than the electroweak scale.

\section{Two infinite classes with rank-3-tensor fermions}

For both these classes the set of fermion multiplets is  

\begin{equation}
\{ {\rm fermion \;\; multiplets} \} \;\;\; =  \;\;\; (\overline{[3]}, \overline{(0)})^a  +  ([2], (1))^b  +  (\overline{[1]}, \overline{(2)})^c 
+ ([0],(3))^d.
\end{equation}

\noindent One class has

\begin{equation}
G= SU(N) \times SU(N-3) \times U(1), \;\;\; (a,b,c,d) = (-(N-3), (N-2), -(N-1), N),
\end{equation}

\noindent while the other class has

\begin{equation}
G= SU(N) \times SU(N-6) \times U(1), \;\;\; (a,b,c,d) = (-(N-6), (N-4), -(N-2), N),
\end{equation}

\noindent Note that these classes show the same basic pattern as the class discussed in section 2. For each fermion multiplet in the set, its ranks under the two non-abelian gauge groups add up to the same number (2 for Eq. (1) and 3 for Eq. (3)). The multiplets are antisymmetric tensors under the larger non-abelian groups and symmetric tensors under the smaller non-abelian group. The signs of the $U(1)$ charges alternate, and their absolute magnitudes are evenly spaced, with the spacing such that the magnitudes of the charges of the multiplets at the end equal the degrees of the two non-abelian groups. 

It is straightforward to check that all six anomalies cancel for the two classes. Again, the fact that the non-abelian anomalies (such as the $SU(N)^3$ and
$SU(N-3)^3$ anomalies for the class shown in Eq. (4)) cancel is non-trivial and involves relationships between the anomalies of multiplets of one groups and the dimensions of the multiplets of the other. The cancellation of $U(1)$ anomalies is also non-trivial, as there are three ratios of $U(1)$ charges, which must satisfy four anomaly conditions, one of which is cubic in the $U(1)$ charges. 

Many of the comments made about the class presented in section 2 apply here also. The presence of symmetric tensors makes it very unlikely that these cases can be embedded in a larger simple group. However, each case is contained in all the cases with larger $N$, from which it can arise by spontaneous symmetry breaking. For the class defined by Eq. (4), the anomaly-free set shown in Eq. (3), when decomposed under the Standard Model subgroup, gives {\it three} families plus extra vectorlike multiplets.
For the class defined by Eq. (5), the anomaly-free set shown in Eq. (3), when decomposed under the Standard Model subgroup, is vectorlike and gives {\it zero} net families ({\it i.e.}, the number of families minus anti-families).

For concreteness  we show the cases with lowest $N$ for the class defined by Eq. (4) in Table III.

\noindent {\bf Table III:} The $N=4,5$, and $6$ cases of the infinite class based on $SU(N) \times SU(N-3) \times U(1)$.
\[
\begin{tabular}{lll}
{\bf N} & {\bf gauge group} & {\bf fermion multiplets} \\
\hline 
$4$ & $SU(4) \times U(1)$  & ${\bf 4}^{-1} + {\bf 6}^{2} + \overline{{\bf 4}}^{-3} + {\bf 1}^4$ \\ & & \\
$5$ & $SU(5) \times SU(2) \times U(1)$ & $({\bf 10}, {\bf 1})^{-2} + ({\bf 10}, {\bf 2})^3 + (\overline{{\bf 5}}, \overline{\bf 3})^{-4} + ({\bf 1}, {\bf 4})^5$ \\ & & \\
$6$ & $SU(6) \times SU(3) \times U(1)$ & $({\bf 20}, {\bf 1})^{-3} + 
({\bf 15}, {\bf 3})^4 + (\overline{{\bf 6}}, \overline{{\bf 6}})^{-5} + ({\bf 1}, {\bf 10})^6$
\end{tabular}
\]

\noindent In the $N=5$ case in Table III, we see that the three Standard Model families of fermions differ both in their $SU(2)$ and $U(1)$ quantum numbers. For example, the three
$\overline{{\bf 5}}$ multiplets form a triplet under the $SU(2)$, whereas the ${\bf 10}$ multiplets form a doublet plus a singlet. Thus the $SU(2) \times U(1)$ could be responsible for non-trivial flavor structure of the Standard Model mass matrices. The same remarks apply to the cases with $N>5$.

The classes presented so far are not the only ones of this type.
It can straightforwardly be seen that there are two infinite anomaly-free classes with 
fermions in the multiplets
 
\begin{equation}
\{ {\rm fermion \;\; multiplets} \} = ([4], (0))^a  +  (\overline{[3]}, \overline{(1)})^b  +  ([2], (2))^c 
+ (\overline{[1]},\overline{(3)})^d + ([0], (4))^e,
\end{equation}

\noindent for one of which the gauge group and $U(1)$ charge assignments given by

\begin{equation}
G= SU(N) \times SU(N-4) \times U(1), \;\;\; (a,b,c,d,e) = ((N-4), -(N-3), (N-2), -(N-1), N),
\end{equation}

\noindent while for the other class they are given by

\begin{equation}
G= SU(N) \times SU(N-8) \times U(1), \;\;\; (a,b,c,d,e) = ((N-8), -(N-6), (N-4), -(N-2), N),
\end{equation}

\noindent For the class defined by Eq. (7), the anomaly-free set shown in Eq. (6), when decomposed under the Standard Model subgroup, gives {\it two} families plus extra vectorlike multiplets.
For the class defined by Eq. (8), the anomaly-free set shown in Eq. (6), when decomposed under the Standard Model subgroup, is vectorlike and gives {\it zero} net families.

\section{Conclusions}
 
The anomaly-free classes pointed out in this paper may be useful for model-building.
The models would typically have extra leptons, which could be chiral or vectorlike under the electroweak group, depending on the scheme. They would have an extra $U(1)$ gauge interaction, which could be broken near the electroweak scale and under which the Standard Model fermions would have distinctive charges unlike those that arise in typical grand unified schemes
\cite{extraU(1)charges}. An interesting feature of many cases (such as the $N=5$ case in Table III) is that there is both a non-abelian and an abelian group that distinguish among the families and therefore act like family symmetries.

The fact that there exist several qualitatively similar infinite classes of cases in which anomalies cancel non-trivially suggests that there is some simple underlying reason, and that there are infinitely many other such infinite classes. However, it does not appear that the underlying reason would be unification in a larger simple group. It would be interesting to discover what the underlying reason is, which would perhaps lead to the discovery of other previously unknown classes where anomalies cancel. Whatever the deeper mathematical or physical reason might be for the existence of these classes, they appear to be interesting from the point of view of model building and phenomenology.

\section*{Acknowledgements} This work was
supported by U.S. DOE under contract DE-FG02-12ER41808.

\end{document}